\documentclass{amsart}
\usepackage{graphicx}
\usepackage{dcolumn}
\usepackage{bm}
\usepackage{amsmath,amsthm,amssymb,stmaryrd}
\usepackage{enumerate}

\begin{document}

\title[Divergent integrals]{Integration by divergent integrals:\\calculus of divergent integrals in term by term integration}

\author{Eric A. Galapon}
\address{Theoretical Physics Group, National Institute of Physics, University of the Philippines, Diliman Quezon City, 1101 Philippines}
\email{eric.galapon@up.edu.ph}
\date{\today}
\maketitle

\begin{abstract}
It is known in the case of the Stieltjes transform that evaluating the integral by expanding the kernel of transformation followed by term by term integration leads to an infinite series of divergent integrals. Moreover, it is known that merely assigning values to the divergent integrals leads to missing terms. This problem of missing terms was earlier resolved in the complex plane, which required the divergent integrals to assume values equal to their finite parts under a set of rules or calculus governing the use of the finite part integral [E.A. Galapon, {\it Proc. Roy. Soc. A} {\bf 473}, 20160567 (2017)]. However, divergent integrals assume a spectrum of values arising from the different ways of removing the divergences in the integration. Here we show that the other interpretations of divergent integrals follow the same set of rules as those of the finite part integral, and then formulate the calculus of divergent integrals arising from term-by-term integration. We find that the calculus does not prescribe a specific interpretation to the divergent integrals but allows us to assign arbitrary values to them, with the missing terms automatically emerging from the chosen interpretation of the divergent integrals itself.
\end{abstract}

\section{Introduction}
When divergent integrals appear in applications, the typical problem is how to tame the divergences in order to assign meaningful values to them \cite{shiekh,lee,blanchet}. However, it is the nature of divergent integrals that they take on a spectrum of values so that no unique value can be assigned to a divergent integral. The different values arise from distinct ways of interpreting divergent integrals corresponding to specific means of removing the divergences in the integration, such as by analytic continuation \cite{analytic}, by regularization \cite{zozulya}, by summability methods \cite{borwein}, and by finite part integrals \cite{monegato,hadamard}, among others \cite{caesaro,distribution,others1,others2,others3}. The different values lead to different results for the target quantity to calculate and ambiguity inevitably arises as to which is the correct one \cite{philips,mitra}. At most only one of the assignable values is correct, if at all. An established fact is that a mere replacement of divergent integrals with values using, say, analytic continuation may yield a result that reproduces the correct result up to some missing terms, a problem known as the problem of missing terms \cite{wong,mcwong,galapon2,galapon3}. It is possible that every interpretation of a divergent integral will always lead to missing terms so that no interpretation could in fact yield the correct result; on the other hand, it is also possible that there is an interpretation that could always yield the exact result; or that which interpretation is appropriate could be decided on a case to case basis. However, at the moment there exists no coherent theory of divergent integrals that can tell us which possibility is in actuality. 

In this paper, we wish to take another step towards solving the problem of missing terms that may, hopefully, point us the way to the development of a rigorous theory of divergent integrals. Recently we made a progress in addressing the problem in the context of evaluating the Stieltjes transform $\int_0^{\infty} f(x) (\omega + x)^{-1}\mathrm{d}x$ for $\omega>0$ \cite{galapon2}. This is a problem where a unique solution is expected for locally integrable $f(x)$ in the entire real line. Divergent integrals arise from this otherwise well defined integral when one attempts to evaluate it by binomially expanding the kernel $(\omega+x)^{-1}$ about $\omega=0$ and then performing a term by term integration. Explicitly, the attempt yields the infinite series $\sum_{k=0}^{\infty} (-1)^k \omega^k \int_0^{\infty} f(x) x^{-k-1}\mathrm{d}x$. If $f(x)$, say, does not vanish at the origin, then each term of the series is a divergent integral. A further attempt at making sense of the expansion by assigning values to the divergent integrals by analytic continuation leads to an answer that reproduces a group of terms of the correct result but completely misses out another group of terms. The fact that analytic continuation yields a group of correct terms intimates that there is some measure of ``mathematical correctness'' to assigning values to the divergent integrals where they appear. But the presence of the missed out terms indicates that assigning values to divergent integrals is just half of the story; the other half is obtaining the missing terms. In \cite{galapon2} we have shown how to rigorously use divergent integrals in evaluating the Stieltjes transform, in particular, how to correctly perform the term-by-term integration and reproduce in the process all the missing terms.

Our solution to the problem involves four key elements. The first element
is the assignment of values to the divergent integrals, $\int_0^{\infty} f(x) x^{-k-1}\mathrm{d}x$, equal to their finite parts. The finite part of $\int_0^{\infty} f(x) x^{-k-1}\mathrm{d}x$ is, by definition, obtained by temporarily excluding the singular point $x=0$ by replacing the lower limit of integration with some $\epsilon>0$ and then dropping all terms that diverge as $\epsilon\rightarrow 0$. The second element is the representation of the finite part of the divergent integrals as a contour integral in the complex plane. This means that the finite part is in fact the value of an absolutely convergent integral, lifting the dubious character of the finite part integral and endowing it the character of a legitimate integral. The third element is the representation of the Stieltjes transform as a contour integral up to a term in the complex plane, with the contour of integration dictated by the contour of integration of the complex integral representation of the finite part integral. The fourth element is the implementation of the term by term integration which involves appropriately choosing the contours of integration in the representation of the Stieltjes transform to satisfy uniform convergence which is lacking in the original term by term integration. From the last element, the interpretation of the divergent integrals as finite part integrals arises naturally and the missing terms emerge from the singularities of the function $f(z) (\omega + z)^{-1}$ in the complex plane. The whole process can be seen as a means of evaluating convergent integrals using divergent integrals, an integration method which we have referred to as finite-part integration. 

The most important lesson that we have learned from finite-part integration is that divergent integrals, at least their finite-parts, are legitimate mathematical objects that can be manipulated to solve well-defined problems \cite{galapon2,galapon3}. 
That is the finite part of a divergent integral can be rigorously employed provided the set of rules governing its use or its calculus is known. The four elements discussed above constitute the calculus of the finite part integral in the evaluation of the Stieltjes transform. However, the finite-part is just one of the potentially infinitely many possible values that can be assigned to the divergent integrals. It is natural and important to ask whether the other assignable values have a calculus of their own that can be applied as a means of evaluating convergent integrals in the same way that the finite integral has a calculus of its own that allows us to use it as a legitimate analytical object in evaluating the Stieltjes transform \cite{galapon2,galapon3}. In this paper, we will show that the same calculus holds for different interpretations of divergent integrals arising from term by term integration. 

As a concrete problem, we will consider in detail the evaluation of the generalized Stieltjes transform
\begin{equation}\label{transform}
	\tilde{S}[f]=\int_{-\infty}^{\infty} \frac{f(x)}{\omega^{2} + x^{2} }\, \mathrm{d}x, \;\;\; \omega>0 ,
\end{equation}
about $\omega =0$ from which the asymptotic behavior of $\tilde{S}[f]$ for small parameters $\omega$ can be directly extracted. This problem is relevant in effective diffusivity in passive advection by laminar and turbulent flow \cite{avellada}, and in non-equilibrium statistical mechanics of self-diffusion as described by the Green-Kubo formula \cite{pavliotis}. However, it is not our objective here to study these problems. Instead, it is our objective to demonstrate by means of equation \eqref{transform} how the assignment of different values to divergent integrals lead to the same correct answer. Divergent integrals arise from equations \eqref{transform}  when we attempt to evaluate the Stieltjes integral by expanding the kernel of the transformation about $\omega=0$, and then performing term by term integration, i.e.,
\begin{equation}\label{termbyterm}
\sum_{k=0}^{\infty} (-1)^k \omega^{2  k} \int_{-\infty}^{\infty} \frac{f(x)}{x^{2 k+2 }}\, \mathrm{d}x,
\end{equation}
which is an infinite series of divergent integrals when $f(x)$ is analytic at the origin. 

We will show how the Stieltjes transform \eqref{transform} can be evaluated by assigning (three) different interpretations to the divergent integrals $\int_{-\infty}^{\infty} f(x) x^{-2k-2}\mathrm{d}x$. In particular, we will show, under certain analyticity condition on $f(x)$, that the transform assumes the evaluation
\begin{equation}\label{solution}
\int_{-\infty}^{\infty}\frac{f(x)}{\omega^2+x^2}\mathrm{d}x = \sum_{k=0}^{\infty} (-1)^k \omega^{2  k} \#\!\!\int_{-\infty}^{\infty} \frac{f(x)}{x^{2 k+2 }}\, \mathrm{d}x + \Delta^{\#}(\omega),
\end{equation}
where $\#\!\int_{-\infty}^{\infty} f(x) x^{-2k-2}\mathrm{d}x$ denotes a particular value assigned to the divergent integral $\int_{-\infty}^{\infty} f(x) x^{-2k-2}\mathrm{d}x$; and $\Delta^{\#}(\omega)$ constitutes the contributions coming from the singularities of the function $f(z)(\omega^2+z^2)^{-1}$ in the complex plane, the contributing singularities being determined by the contour integral representation of the chosen interpretation of the divergent integrals. The first term of equation \eqref{solution} is the result of term-by-term integration with a particular value assigned to the divergent integrals; and the second term represents the group of terms missed out by naive term-by-term integration. From our example, we will proceed to formulate explicitly the calculus of divergent integrals arising from term-by-term integration. Our main conclusion is that the calculus does not prescribe a specific interpretation to the divergent integrals arising from term by term integration but allows us to assign arbitrary values to them, with the missing terms automatically emerging from the chosen interpretation of the divergent integrals itself. 

The paper is organized as follows. In Section-\ref{manyvalues} we discuss two classes of interpretations of divergent integrals. In Section-\ref{cauchyfox} we assign three different values to divergent integrals whose divergences arise from a non-integrable singularity in the interior of the interval of integration, a divergence we refer to as Cauchy-Fox singularity \cite{fox}; the divergent integrals in expression \eqref{termbyterm} fall under this category. In Section-\ref{representations} we obtain the contour integral representations of the three assigned values to Cauchy-Fox divergent integrals. In Section-\ref{howto} we evaluate the generalized Stieltjes transform using the three different values of the divergent integrals in expression \eqref{termbyterm}. In Section-\ref{hilbertransform} we evaluate the Hilbert transform to address the question of the possibility of term by term integration without missing out contributing terms. In Section-\ref{calculus} we formulate the calculus of divergent integrals arising from term-by-term integration. In Section-\ref{application} we apply the calculus of divergent integrals to the evaluation of Fourier transforms of products of exponential type functions. In Section-\ref{conclusion} we conclude.

\section{The many values of divergent integrals} \label{manyvalues}
Suppose that the following integral is divergent due to some non-integrable singularity of the integrand in the 
interval of integration,
\begin{equation}
\int_a^b h(x)\, \mathrm{d}x .
\end{equation}
A value is assigned to this divergent integral by a systematic removal of the offending singularity from which a finite value is extracted and interpreted as a value of the divergent integral. However, there is no unique way of removing the singularity and there are several competing ways of doing so. The only fundamental restriction on the possible assignable values is that they should reproduce the unique value of the integral when the integral happens to be absolutely convergent. Here we will restrict ourselves to two ways. One is by an appropriate modification of the integrand and another is by an appropriate modification of the limits of integration. 

In the former, the integrand is modified by way of a sequence of functions $\{h_{\nu}\}$ with the property that $\lim_{\nu\rightarrow \infty}h_{\nu}(x)=h(x)$.  Then one assigns the value to the divergent integral
\begin{equation}
\#\!\!\int_a^b h(x)\,\mathrm{d}x = \lim_{\nu\rightarrow \infty}\int_a^b h_{\nu}(x)\, \mathrm{d}x,
\end{equation}
provided the limit exists. This method falls under the summability method \cite{borwein}. Clearly there are potentially infinitely many possible values that can be assigned to the divergent integral corresponding to the infinitely many possible sequences of functions converging to $h(x)$. This way of interpreting the divergent integral is well motivated in physics because many of the divergences arises from some haphazard interchange of orders of operations without regard to the necessary uniformity conditions that may allow the interchange. In fact, this is routinely employed in interpreting scattering Green's functions assuming divergent integral representations.

On the other hand, the later involves deleting the contribution of the offending non-integrable singularity by excluding the singularity from the integration. Suppose $x_0$ is a non-integrable singularity of $h(x)$ in the interior of the interval $[a,b]$. Then $x_0$ can be excluded from the integration in the following way,
\begin{equation}\label{modified2}
\int_a^{x_0-\epsilon_1} h(x)\, \mathrm{d}x + \int_{x_0+\epsilon_2}^b h(x)\, \mathrm{d}x
\end{equation}
for some $\epsilon_1, \epsilon_2>0$ which are not necessarily equal. Formally, 
expression \eqref{modified2} approaches the divergent integral as $\epsilon_1,\epsilon_2\rightarrow 0$; however, it can be written in the form
\begin{equation}
\int_a^{x_0-\epsilon_1} h(x)\, \mathrm{d}x + \int_{x_0+\epsilon_2}^b h(x)\, \mathrm{d}x = C(\epsilon_1,\epsilon_2) + D(\epsilon_1,\epsilon_2)
\end{equation}
where $C(\epsilon_1,\epsilon_2)$ comprises the group of terms that converge as $\epsilon_1,\epsilon_2\rightarrow 0$ and $D(\epsilon_1,\epsilon_2)$ comprises the terms that diverge as in the same limit. Then a value is assigned to the divergent integral by assigning it the value equal to the limit of the converging part, 
\begin{equation}\label{value1}
\#\!\!\int_a^b h(x)\, \mathrm{d}x = \lim_{\epsilon_1,\epsilon_2\rightarrow 0}. C(\epsilon_1,\epsilon_2) .
\end{equation}
Equivalently, this value can be obtained by dropping the diverging part outrightly and then taking the limit $\epsilon_1,\epsilon_2\rightarrow 0$, that is,
\begin{equation}\label{value2}
\#\!\!\int_a^b h(x)\, \mathrm{d}x = \lim_{\epsilon_1,\epsilon_2\rightarrow 0} \left(\int_a^{x_0-\epsilon_1} h(x)\, \mathrm{d}x + \int_{x_0+\epsilon_2}^b h(x)\, \mathrm{d}x - D(\epsilon_1,\epsilon_2)\right) .
\end{equation}
Expressions \eqref{value1} and \eqref{value2} are equivalent. When $\epsilon_1=\epsilon_2$, the value is known as the Hadamard finite part or simply the finite part of the divergent integral \cite{monegato}. A value different from the finite part can be obtained when $\epsilon_1$ and $\epsilon_2$ are not independent of each other, say $\epsilon_2=c\epsilon_1$ where $c$ is some constant different from $1$. The value will depend on $c$, which is arbitrary. Again deletion of the singularity leads to many possible values depending on the relationship between the $\epsilon$'s.

\section{Divergent integrals arising from Cauchy-Fox type singularities}\label{cauchyfox}

We now consider the problem of assigning values to a family of divergent integrals arising from a non-integrable singularity in the interior of the interval of integration. The family consists of the divergent integrals
\begin{equation}\label{divergent}
\int_a^b \frac{f(x)}{(x-x_0)^{n+1}}\, \mathrm{d}x, \;\;\; a < x_0 < b , \;\;\; n=0, 1, 2, \dots ,
\end{equation}
where $f(x)$ is locally integrable in the entire interval of integration. The divergent integrals in expression \eqref{termbyterm} belong to this family. There are potentially infinitely many possible values that we can assign to these integrals. We denote one such value by $\#\!\!\int_a^b  f(x)(x-x_0)^{-n-1}\, \mathrm{d}x$, where $\#$ serves to identify the assigned value or a specific interpretation of the divergent integral \eqref{divergent}. 

We will assume that the limits of integration, $a$ and $b$, are both finite; but when one, say, the upper limit of integration is infinite, a value, $\#\!\!\int_a^{\infty} f(x)(x-x_0)^{-n-1}\mathrm{d}x$, is assigned as follows, 
\begin{equation}
\#\!\!\int_a^{\infty} \frac{f(x)}{ (x-x_0)^{n+1}}\mathrm{d}x = \lim_{b\rightarrow\infty}\#\!\!\int_a^{b}\frac{f(x)}{ (x-x_0)^{n+1}}\mathrm{d}x, \;\;\; a<x_0<b<\infty,
\end{equation}
provided the limit exists. The same interpretation is meant when the lower limit is infinite. When both limits are infinite, we will mean the assignment 
\begin{equation}
\#\!\!\!\int_{-\infty}^{\infty} \frac{f(x)}{(x-x_0)^{n+1}}\,\mathrm{d}x = \lim_{c\rightarrow\infty} \int_{-c}^{c} \frac{f(x)}{(x-x_0)^{n+1}}\,\mathrm{d}x, \;\; -c<x_0<c .
\end{equation}
Hence it will be sufficient for us to consider the case when both limits are finite. 

We now assign a value by summability method. A value is assigned by avoiding the non-integrable singularity at $x=x_0$ by a judicious choice of a sequence of functions that converge to the integrand with each function in the sequence satisfying local integrability in the entire interval of integration. There are infinitely many possible such sequences. Here we chose the pair of sequences $f(x)(x-(x_0\pm i \epsilon))^{-n-1}$, for $\epsilon>0$; clearly both sequences converge to the singular integrand $f(x)(x-x_0)^{-n-1}$ as $\epsilon\rightarrow 0$. We then assign the  corresponding values
\begin{equation}
\mathrm{UBV}\!\!\int_a^b \frac{f(x)}{(x-x_0)^{n+1}}\, \mathrm{d}x = \lim_{\epsilon\rightarrow 0} \int_a^b \frac{f(x)}{(x-(x_0+i \epsilon))^{n+1}}\, \mathrm{d}x ,
\end{equation}
\begin{equation}
\mathrm{LBV}\!\!\int_a^b \frac{f(x)}{(x-x_0)^{n+1}}\, \mathrm{d}x = \lim_{\epsilon\rightarrow 0} \int_a^b \frac{f(x)}{(x-(x_0-i \epsilon))^{n+1}}\, \mathrm{d}x ,
\end{equation}
provided the limits exist. UBV is the acronym for upper-boundary-value; and LBV is for lower-boundary-value. Our reference to them as upper and lower boundary values follows from the fact that they are specific values of the complex valued function due to Fox \cite{fox},
\begin{equation}
\Phi(z)=\int_a^b \frac{f(x)}{(x-z)^{n+1}}\, \mathrm{d}x, \;\;\; z\in \mathbb{C}\backslash [a,b] .
\end{equation}
In particular, the UBV is the value of $\Phi(z)$ as $z\rightarrow x_0$ from above; and the LBV is the value as $z\rightarrow x_0$ from below. By Fox's results, the boundary values exist \cite{fox}. 

Another value that can be assigned to the divergent integrals is obtained by a symmetric removal of the offending singularity and eventually dropping the terms that diverge in the limit. This value is the finite part of the divergent integral and are given  by the Cauchy and Fox principal values \cite{fox}, 
\begin{eqnarray}\label{fox}
&&\mathrm{FPI}\!\!\int_a^b \frac{f(x)}{(x-x_0)^{n+1}} \mbox{d}x \nonumber \\
&&\hspace{12mm}=\lim_{\epsilon\rightarrow 0^+}\left[\int_a^{x_0-\epsilon} \frac{f(x)}{(x-x_0)^{n+1}} \mathrm{d}x+ \int_{x_0+\epsilon}^{b} \frac{f(x)}{(x-x_0)^{n+1}} \mathrm{d}x - H_n(x_0,\epsilon)\right],
\end{eqnarray}
where
\begin{eqnarray}
H_n(x_0,\epsilon)&=&0, \;\;\; n=0\label{fox1} \\
H_n(x_0,\epsilon)&=& \sum_{k=0}^{n-1} \frac{f^{(k)}(x_0)}{k!(n-k)} \frac{(1-(-1)^{n-k})}{\epsilon^{n-k}},\;\;\; n=1, 2, \dots  , \label{fox2}
\end{eqnarray}
The $n=0$ case is the well-known Cauchy principal value. The term that is subtracted, $H_n(x_0,\epsilon)$, is precisely the term that diverges in the limit $\epsilon\rightarrow 0$.

In general, these three values will not be equal. However, they are not unrelated. Fox established much earlier the relationship that the finite part integral is just the simple average of the two boundary values \cite{fox}. Below we will derive the relationship among these values under certain analyticity condition. 

\section{Contour integral representations of the LBV, UBV and FPI}\label{representations}
We now establish the complex contour integral representations of the UBV, LBV and FPI of the divergent integral $\int_a^b f(x) (x-x_0)^{-n-1}\mathrm{d}x$. 
This requires some analyticity assumption on the function $f(x)$ when extended in the complex plane.  Given $f(x)$ we construct the function $f(z)$ which is obtained from $f(x)$ by replacing the real variable $x$ with the complex variable $z$; that is, $f(z)$ is the complex extension of $f(x)$.  We assume all throughout that $f(z)$ is analytic in the closed interval $[a,b]$; this uniquely defines $f(z)$ in the entire complex plane by analytic continuation. Then there is some open region $R$ containing the interval $[a,b]$ such that $f(z)$ is analytic in the entire $R$. 

\begin{figure}
	\includegraphics[scale=0.5]{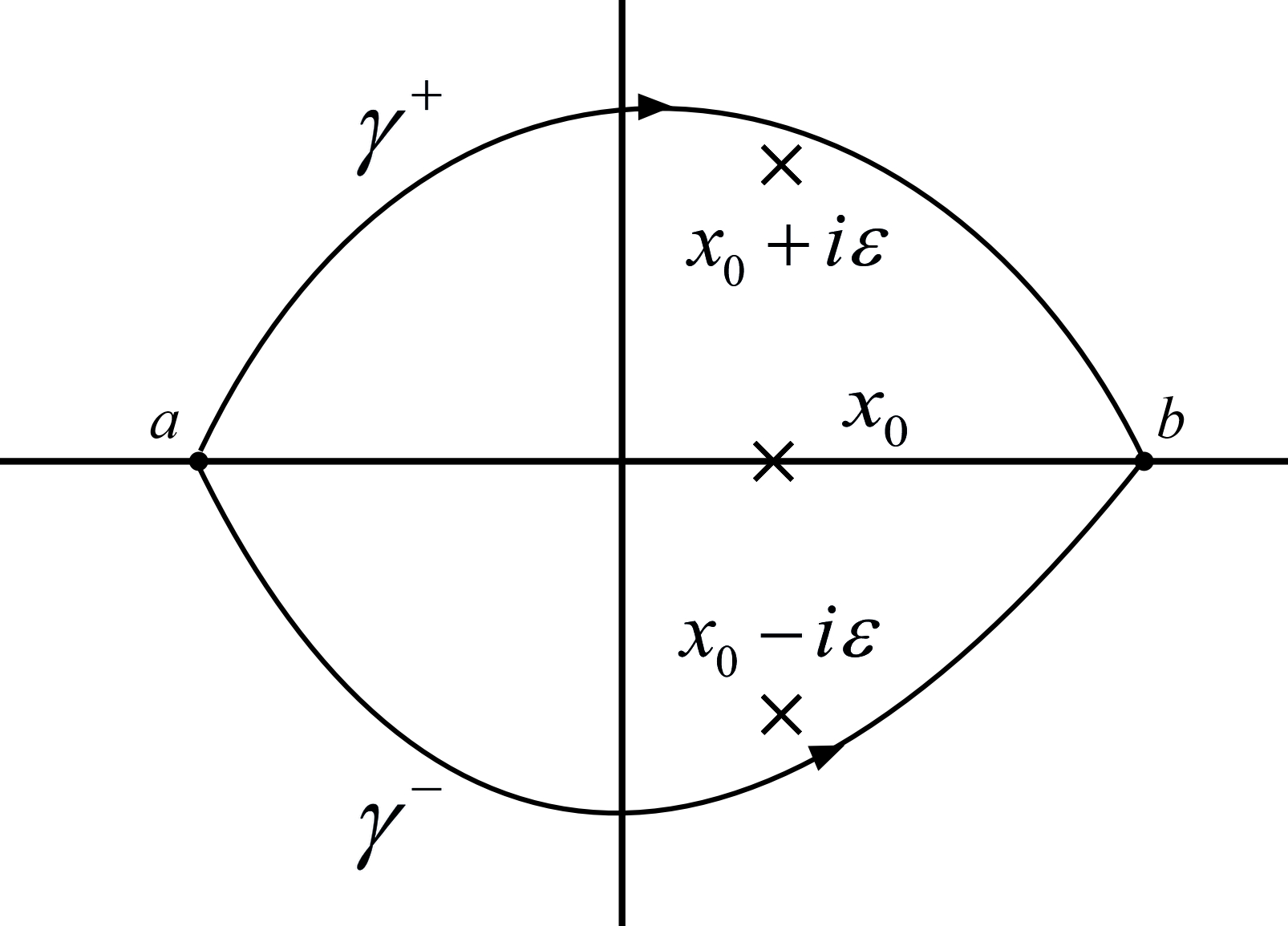}
	\caption{The paths of integration in the contour integral representations of the UBV, LBV and FPI of the divergent integral $\int_a^b f(x) (x-x_0)^{-n-1}\,\mathrm{d}x$. }
	\label{fig:boat1}
\end{figure}

Let $\gamma^+$ be a simple path in $R$ starting at the point $a$ and ending at $b$ that passes through the upper half plane; also let $\gamma^-$ be a simple path in $R$ starting at the point $a$ and ending at $b$ that passes through the lower half plane; both paths are depicted in Figure-1. Any singularity of $f(z)$ lies outside of the region bounded by $\gamma^+$ and $\gamma^-$. For some sufficiently small $\epsilon>0$, such that $(x_0+i\epsilon)$ is enclosed by $\gamma^+$ and the interval $[a,b]$, the function $f(z)/(z-(x_0+i\epsilon))^{n+1}$ is analytic everywhere in $R$ except at $z=x_0+i\epsilon$, which is a pole of order $(n+1)$. Let $C$ be the closed contour comprising the straight line from $a$ to $b$ and the path $\gamma^+$ traversed in the opposite direction. Since $C$ encloses the pole $(x_0+i\epsilon)$, we have by the residue theorem
\begin{eqnarray}
\oint_C \frac{f(z)}{\left(z-(x_0+i \epsilon)\right)^{n+1}}\,  \mathrm{d}z = 2\pi i \frac{f^{(n)}(x_0+i\epsilon)}{n!} .
\end{eqnarray}
On decomposing the contour in the left hand side of the equation along the straight path $[a,b]$ and the path through $\gamma^+$ and taking the limit $\epsilon\rightarrow 0$, we arrive at 
\begin{eqnarray}
\lim_{\epsilon\rightarrow 0}\int_a^b \frac{f(x)}{(x-(x_0+i \epsilon))^{n+1}} \, \mathrm{d}x &=& \lim_{\epsilon\rightarrow 0} \int_{\gamma^+} \frac{f(z)}{\left(z-(x_0+i \epsilon)\right)^{n+1}}\,  \mathrm{d}z \nonumber \\
&& \hspace{12mm} + \lim_{\epsilon\rightarrow 0} 2\pi i \frac{f^{(n)}(x_0+i\epsilon)}{n!}.
\end{eqnarray}
The left hand side is just the upper boundary value UBV. While the limit and the integral cannot be interchanged in the left hand side because the integral will fail to converge under the interchange, the limit and the integral can be interchanged in the first term of the right hand side of the equation. Then we have
\begin{equation}
\mathrm{UBV}\!\!\int_a^b \frac{f(x)}{(x-x_0)^{n+1}} \,\mathrm{d}x=\int_{\gamma^+} \frac{f(z)}{(z-x_0)^{n+1}}\, \mathrm{d}z + 2\pi i \frac{f^{(n)}(x_0)}{n!}.\label{pos1}
\end{equation}
On the other hand, if we close the contour below via the path $\gamma^-$ and perform the same limiting procedure, we obtain
\begin{equation}
\mathrm{UBV}\!\!\int_a^b \frac{f(x)}{(x-x_0)^{n+1}} \,\mathrm{d}x=\int_{\gamma^-} \frac{f(z)}{(z-x_0)^{n+1}}\, \mathrm{d}z\label{pos2},
\end{equation}
where there is no residue term because the pole is outside the contour. Equation \eqref{pos2} shows that the UBV is the value of an absolutely convergent integral, and it gives the desired complex contour integral representation of the UBV.

Also let us consider the function $f(z)/(z-(x_0-i\epsilon))^{n+1}$, still for sufficiently small $\epsilon>0$ such that $(x_0-i\epsilon)$ is enclosed by $\gamma^-$ and the interval $[a,b]$. This time we have a pole below the real axis at $z=x_0-i\epsilon$ of order $n+1$ as well. Closing the contour below the real axis through $\gamma^-$ and evaluating the limit $\epsilon\rightarrow 0$, we obtain the lower boundary value in terms of the contour $\gamma^-$,
\begin{equation}
\mathrm{LBV}\!\!\int_a^b \frac{f(x)}{(x-x_0)^{n+1}} \,\mathrm{d}x=\int_{\gamma^-} \frac{f(z)}{(z-x_0)^{n+1}}\, \mathrm{d}z - 2\pi i \frac{f^{(n)}(x_0)}{n!}.\label{neg2}
\end{equation}
On the other hand, closing the contour above the real axis and performing the same limit, we obtain the lower boundary value in terms of the contour $\gamma^+$,
\begin{equation}
\mathrm{LBV}\!\!\int_a^b \frac{f(x)}{(x-x_0)^{n+1}} \,\mathrm{d}x=\int_{\gamma^+} \frac{f(z)}{(z-x_0)^{n+1}}\, \mathrm{d}z . \label{neg1}
\end{equation}
Equation \eqref{neg1} similarly shows that the LBV is the value of an absolutely convergent integral, and it gives the desired contour integral representation of the LBV.

On the other hand, it is established in \cite{galapon1} that the finite part, under the same condition of analyticity on $f(z)$, is the average of the contour integrals along $\gamma^+$ and $\gamma^-$,
\begin{equation}\label{fpi}
\mathrm{FPI}\!\!\int_a^b \frac{f(x)}{(x-x_0)^{n+1}} \,\mathrm{d}x=\frac{1}{2}\left(\int_{\gamma^+} \frac{f(x)}{(x-x_0)^{n+1}} \,\mathrm{d}x+\int_{\gamma^-} \frac{f(z)}{(z-x_0)^{n+1}}\, \mathrm{d}z\right).
\end{equation}
This result is consistent with Fox's result that the finite part is the simple average of the boundary values of the function $\Phi(z)$ \cite{fox}. Again \eqref{fpi} shows that the FPI is the value of an absolutely convergent integral, and it provides the contour integral representation of the FPI.

From the preceding representations of the three values assigned to the divergent integrals, we can readily establish the relationships among them. From equations \eqref{pos1} and \eqref{neg1}, we obtain the difference between the boundary values,
\begin{equation}\label{bvdifference}
\mathrm{UBV}\!\!\int_a^b \frac{f(x)}{(x-x_0)^{n+1}} \,\mathrm{d}x-\mathrm{LBV}\!\!\int_a^b \frac{f(x)}{(z-x_0)^{n+1}}\, \mathrm{d}x = 2\pi i \frac{f^{(n)}(x_0)}{n!}.
\end{equation} 
On the other hand, from equation \eqref{fpi} and the contour integral representations of the boundary values, we have the relationship
\begin{equation}
\mathrm{FPI}\!\!\int_a^b \frac{f(x)}{(x-x_0)^{n+1}} \,\mathrm{d}x=\frac{1}{2}\left(\mathrm{UBV}\!\!\int_a^b \frac{f(x)}{(x-x_0)^{n+1}} \,\mathrm{d}x+\mathrm{LBV}\!\!\int_a^b \frac{f(x)}{(x-x_0)^{n+1}}\, \mathrm{d}x\right)
\end{equation}
This, together with equation \eqref{bvdifference}, yields the relationship of the finite part integral and either of the boundary values,
\begin{equation}\label{fpiubv}
\mathrm{FPI}\!\!\int_a^b \frac{f(x)}{(x-x_0)^{n+1}} \,\mathrm{d}x=\mathrm{UBV}\!\!\int_a^b \frac{f(x)}{(x-x_0)^{n+1}} \,\mathrm{d}x - \pi i \frac{f^{(n)}(x_0)}{n!},
\end{equation}
\begin{equation}\label{fpilbv}
\mathrm{FPI}\!\!\int_a^b \frac{f(x)}{(x-x_0)^{n+1}} \,\mathrm{d}x=\mathrm{LBV}\!\!\int_a^b \frac{f(x)}{(x-x_0)^{n+1}} \,\mathrm{d}x + \pi i \frac{f^{(n)}(x_0)}{n!} .
\end{equation}
Clearly, when the $n$-th derivative vanishes at the singularity, the three assigned values to the divergent integrals are equal; otherwise, they assign different values.

\section{Integration of the Stieltjes transform by divergent integrals}\label{howto}
We now proceed to demonstrate how the three different interpretations of the divergent integrals $\int_a^b f(x) (x-x_0)^{-n-1}\,\mathrm{d}x$ lead to the exact evaluation of the generalized Stieltjes transform in three different ways. The previous two Sections implement the first two elements of integration using divergent integrals which are assigning an interpretation to the emergent ill-defined integrals, and casting the corresponding value as a complex contour integral. Here we implement the last two elements which are representing the transform as a contour integral along the paths of the integral representation of the intended interpretation of the divergent integrals, and choosing the appropriate contours of representation to allow term by term integration.

\subsection{Complex contour integral representations of $\int_a^b g(x)\mathrm{d}x$}

Let $g(x)$ be locally integrable in the interval $[a,b]$ of the real line so that the following integral exists,  
\begin{equation}\label{convergent}
\int_a^b g(x)\,\mathrm{d}x ,
\end{equation}  
i.e. the integral converges. Depending on the analytic properties of the complex extension, $g(z)$, of $g(x)$, there are, in principle, various complex contour integral representations of the integral \eqref{convergent} arising from various distinct paths in the complex plane connecting $a$ and $b$ that can be continuously deformed into the segment $[a,b]$. From the representations involving single paths of integration, more representations can be constructed by direct addition of two more representations. Here we wish to represent the integral \eqref{convergent} in terms of the paths $\Gamma^+$ and $\Gamma^-$, depicted in Figure-2, which are homotopic to the paths $\gamma^+$ and $\gamma^-$ in the contour integral representations of the LBV, UBV and FPI, under the assumption that $g(x)$ possesses a complex extension, $g(z)$, that is analytic in the interval $[a,b]$ and is at most meromorphic elsewhere. In the development to follow, we require that neither path passes through any pole of $g(z)$. 

Let us consider the integral $\int_{\Gamma^+}g(z)\mathrm{d}z$. We deform the contour $\Gamma^+$ into the contour $\tilde{\Gamma}^+$ as shown in Figure-2. This breaks the integral $\int_{\Gamma^+}g(z)\mathrm{d}z$ into two pieces,
\begin{equation}\label{upper}
\int_{\Gamma^+} g(z)\, \mathrm{d}z = \int_a^b g(x)\ \mathrm{d}x - 2\pi i \sum_k \mathrm{Res}[g(z)]_{z=z_k^+}
\end{equation}
where the $z_k^+$'s are the poles of $g(z)$ that are in the upper half plane and enclosed by $\Gamma^+$ and the interval $[a,b]$. Similarly, we consider the integral $\int_{\Gamma^-} g(z)\mathrm{d}z$, and deform the path $\Gamma^-$ into the contour $\tilde{\Gamma}^-$ as shown again in Figure-2. This, too, breaks the integral $\int_{\Gamma^-} g(z)\mathrm{d}z$ in two parts,
\begin{equation}\label{lower}
\int_{\Gamma^-} g(z)\, \mathrm{d}z = \int_a^b g(x)\ \mathrm{d}x + 2\pi i \sum_k \mathrm{Res}[g(z)]_{z=z_k^-}
\end{equation}
where the $z_k^-$'s are the poles of $g(z)$ that are in the lower half-plane and enclosed by $\Gamma^-$ and the interval $[a,b]$.

From equations \eqref{upper} and \eqref{lower}, we obtain two different contour integral representations of the given integral,
\begin{equation}\label{upperrep}
\int_a^b g(x)\ \mathrm{d}x=\int_{\Gamma^+} g(z)\, \mathrm{d}z + 2\pi i \sum_k \mathrm{Res}[g(z)]_{z=z_k^+},
\end{equation}
\begin{equation}\label{lowerrep}
\int_a^b g(x)\ \mathrm{d}x=\int_{\Gamma^-} g(z)\, \mathrm{d}z - 2\pi i \sum_k \mathrm{Res}[g(z)]_{z=z_k^-}.
\end{equation}
Adding equations \eqref{upperrep}and \eqref{lowerrep} and dividing by two, we obtain another contour integral representation
\begin{eqnarray}
\int_a^b g(x)\,\mathrm{d}x &=& \frac{1}{2} \left(\int_{\Gamma^+} g(z)\, \mathrm{d}z + \int_{\Gamma^-} g(z)\,\mathrm{d}z\right) \nonumber \\
&& \hspace{4mm.} + \pi i \left(\sum_k \mathrm{Res}[g(z)]_{z=z^+_k}-\sum_k \mathrm{Res}[g(z)]_{z=z^-_k}\right) .\label{fpirep}
\end{eqnarray}
These are just three of the many possible complex representations of the given integral, but they are just what we need here.

\begin{figure}
	\includegraphics[scale=0.5]{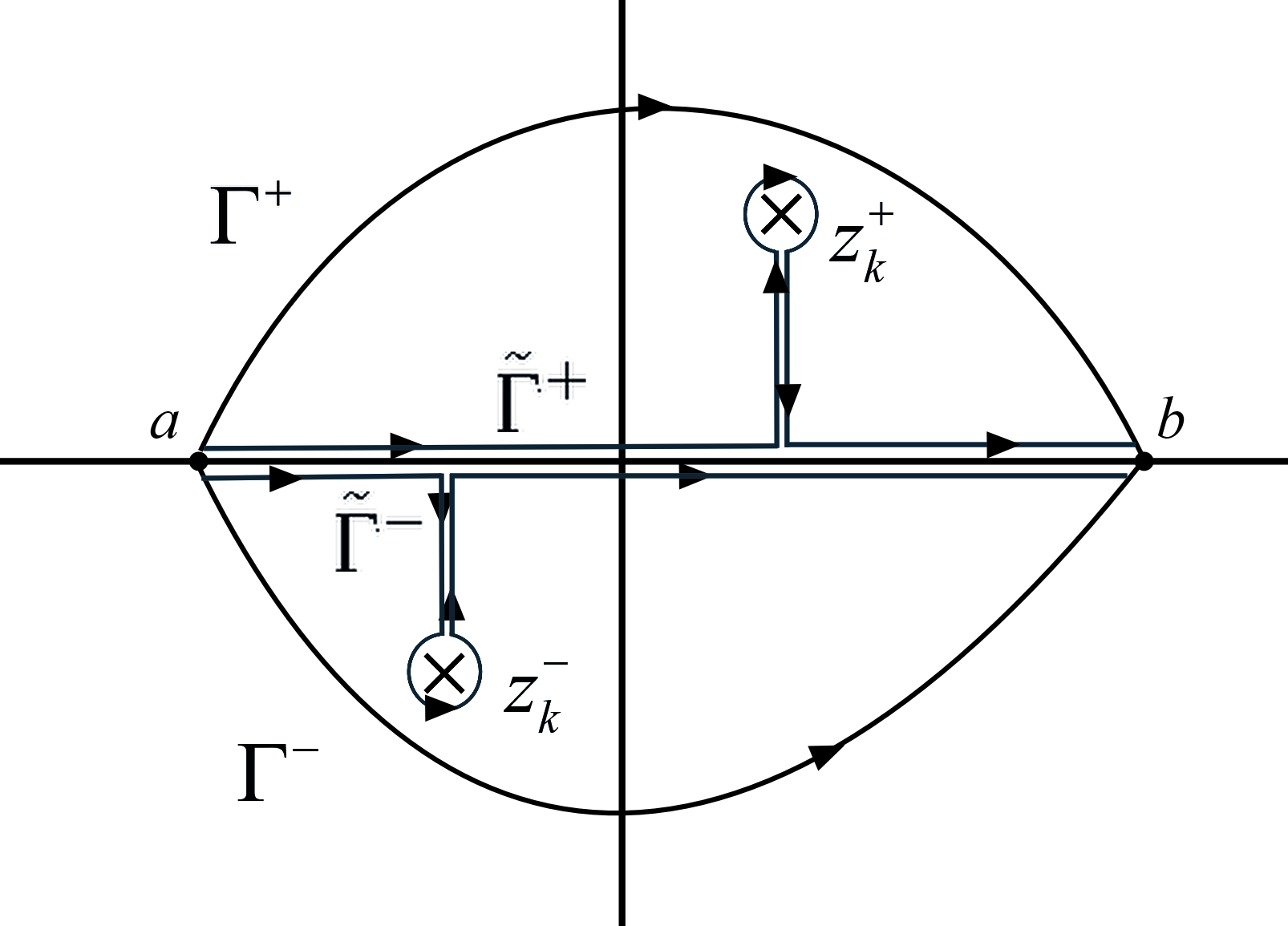}
	\caption{The paths of integration for the contour integral representations of the integral $\int_a^b g(x)\mathrm{d}x$. The $z_k^+$'s are the poles of $g(z)$ in the upper half plane that are below $\Gamma^+$; and the $z_k^-$'s are the poles of $g(z)$ in the lower half plane that are above $\Gamma^-$.}
	\label{fig:boat2}
\end{figure}

\subsection{Integration of the Stieltjes transform}
We now have all the ingredients to evaluate the Stieltjes transform using the three different values of the divergent integral $\int_0^{\infty} f(x) x^{-2k-2}\mathrm{d}x$. We assume that $f(x)$ has the property that the Stieltjes transform converges absolutely. Then we have the equality
\begin{equation}\label{equiv}
\int_{-\infty}^{\infty} \frac{f(x)}{(\omega^2 + x^2)} \, \mathrm{d}x =\lim_{a\rightarrow \infty}\int_{-a}^{a} \frac{f(x)}{(\omega^2 + x^2)} \, \mathrm{d}x .
\end{equation}
We further assume that the complex extension of $f(x)$, $f(z)$, is entire so that the poles of $f(z) (\omega^2 + z^2)^{-1}$ come from the poles of $(\omega^2+z^2)^{-1}$ which are $\pm i \omega$. 

\subsubsection{Integration by the $\mathrm{LBV}$} To evaluate with the LBV, we represent the integral in the right hand side of \eqref{equiv} as a contour integral up to a term using a contour of integration consistent with the contour integral representation of the LBV. We have at least two options. One is the simple contour $\Gamma_1^+$ in the upper half plane that starts at $-a$ and ends at $a$, and that passes below the pole $i\omega$; another is a similar contour $\Gamma_2^+$ in the upper half plane only that it passes above the pole $i\omega$. (See Figure-3.) Both contours can represent the integral in the complex plane, but we will demonstrate that only one of them will allow term by term integration.  

Using the contour $\Gamma_1^+$, the given integral assumes the contour integral representation
\begin{equation}\label{lbvrep0}
\int_{-a}^{a} \frac{f(x)}{(\omega^2 + x^2)} \, \mathrm{d}x = \int_{\Gamma^+} \frac{f(z)}{\omega^2 + z^2}\, \mathrm{d}z,
\end{equation}
where there is no contribution coming from the residue at the pole $i\omega$ because the pole is above the contour $\Gamma_1^+$. Now we expand the kernel binomially,
\begin{equation}\label{binomial}
\frac{1}{\omega^2+z^2} = \sum_{j=0}^{\infty}(-1)^j \frac{\omega^{2j}}{z^{2j+2}}
,
\end{equation}
with the intention to substitute the expansion back into equation \eqref{lbvrep0} and perform term by term integration. Assuming that the integration and summation can be interchanged, we obtain the evaluation
\begin{eqnarray}
\int_{-a}^{a} \frac{f(x)}{(\omega^2 + x^2)} \, \mathrm{d}x &=& \int_{\Gamma^+}\left[\sum_{j=0}^{\infty}(-1)^j \omega^{2j} \frac{f(z)}{z^{2j+2}}\right]\, \mathrm{d}z\label{wrong1}\\
&=& \sum_{j=0}^{\infty}(-1)^j \omega^{2j}\int_{\Gamma^+} \frac{f(z)}{z^{2j+2}}\, \mathrm{d}z\label{wrong2} \\
&=&\sum_{j=0}^{\infty}(-1)^j \omega^{2j}\mathrm{LBV}\!\!\int_{-a}^{a} \frac{f(x)}{x^{2j+2}}\, \mathrm{d}x,\label{wrong}
\end{eqnarray}
where the last line \eqref{wrong} follows from the identification of the integral in line \eqref{wrong2} as the LBV of the divergent integral as given by equation \eqref{neg1}. 
Equation \eqref{wrong} is just the result of term by term integration with the divergent integrals interpreted by their lower boundary values. A notable feature of expression \eqref{wrong} is the absence of contribution coming from the pole at $i\omega$.

However, the integration and summation cannot be interchanged in the LHS of equation \eqref{wrong1} because the expansion does not converge uniformly along the path of integration $\Gamma_1^+$. The expansion \eqref{binomial} converges only for all $z$ satisfying $|z|>\omega$. Since the pole $i\omega$ is above $\Gamma_1^+$ there is inevitably a segment of $\Gamma_1^+$ in which all $z$ there satisfy $|z|<\omega$. In that segment the expansion does not converge and the interchange cannot be implemented. Then the LHS of equation \eqref{wrong2} does not follow from the previous line \eqref{wrong1}, so that expression \eqref{wrong} is not the correct value of the integral. 

Now using instead the path of integration $\Gamma_2^+$ for the contour in equation \eqref{upperrep}, the integral assumes the contour integral representation
\begin{equation}\label{lbvrep}
\int_{-a}^{a} \frac{f(x)}{(\omega^2 + x^2)} \, \mathrm{d}x = \int_{\Gamma_2^+} \frac{f(z)}{\omega^2 + z^2}\, \mathrm{d}z + \frac{\pi}{\omega} f(i\omega),
\end{equation}
where the second term is the contribution coming from the pole at $i\omega$. Again we attempt to substitute the expansion of the kernel back into the integral with the intention to perform term by term integration. As discussed above, this is only possible when $|z|>\omega$ for all $z$ along $\Gamma_2^+$. This immediately imposes the condition $a>\omega$. When this condition is satisfied, we can always deform $\Gamma_2^+$ such that for all $z$ in $\Gamma_2^+$ the inequality $|z|>\omega$ is satisfied; we can in particular deform it into the semi-circle $\Gamma^+$ with radius $a$ as depicted in Figure-2. In this path of integration, the expansion converges uniformly so that term by term integration is allowed; we then obtain the result
\begin{eqnarray}\label{lbvrep4}
\int_{-a}^{a} \frac{f(x)}{(\omega^2 + x^2)} \, \mathrm{d}x &=& \sum_{j=0}^{\infty}(-1)^j \omega^{2n}\int_{\Gamma^+} \frac{f(z)}{z^{2j+2}}\, \mathrm{d}z  + \frac{\pi}{\omega} f(i\omega) . \nonumber \\
&=&\sum_{j=0}^{\infty}(-1)^j \omega^{2n} \; \mathrm{LBV}\!\!\int_{-a}^a \frac{f(x)}{x^{2j+2}}\, \mathrm{d}x +   \frac{\pi}{\omega} f(i\omega), \;\;\; \omega<a .
\end{eqnarray}
(An alternative but less intuitive way of arriving at this result is given in the Supplementary material.)

On its own, equation \eqref{lbvrep4}  is already the desired result of the integration  for some fixed finite $a$. Clearly the first term is the result of term by term integration of the binomial expansion of the kernel with the divergent integrals interpreted as their lower boundary values. On the other hand, the second term is the missed term by naive term by term integration followed by a mere assignment of the divergent integrals with their lower boundary values.

\begin{figure}
	\includegraphics[scale=0.5]{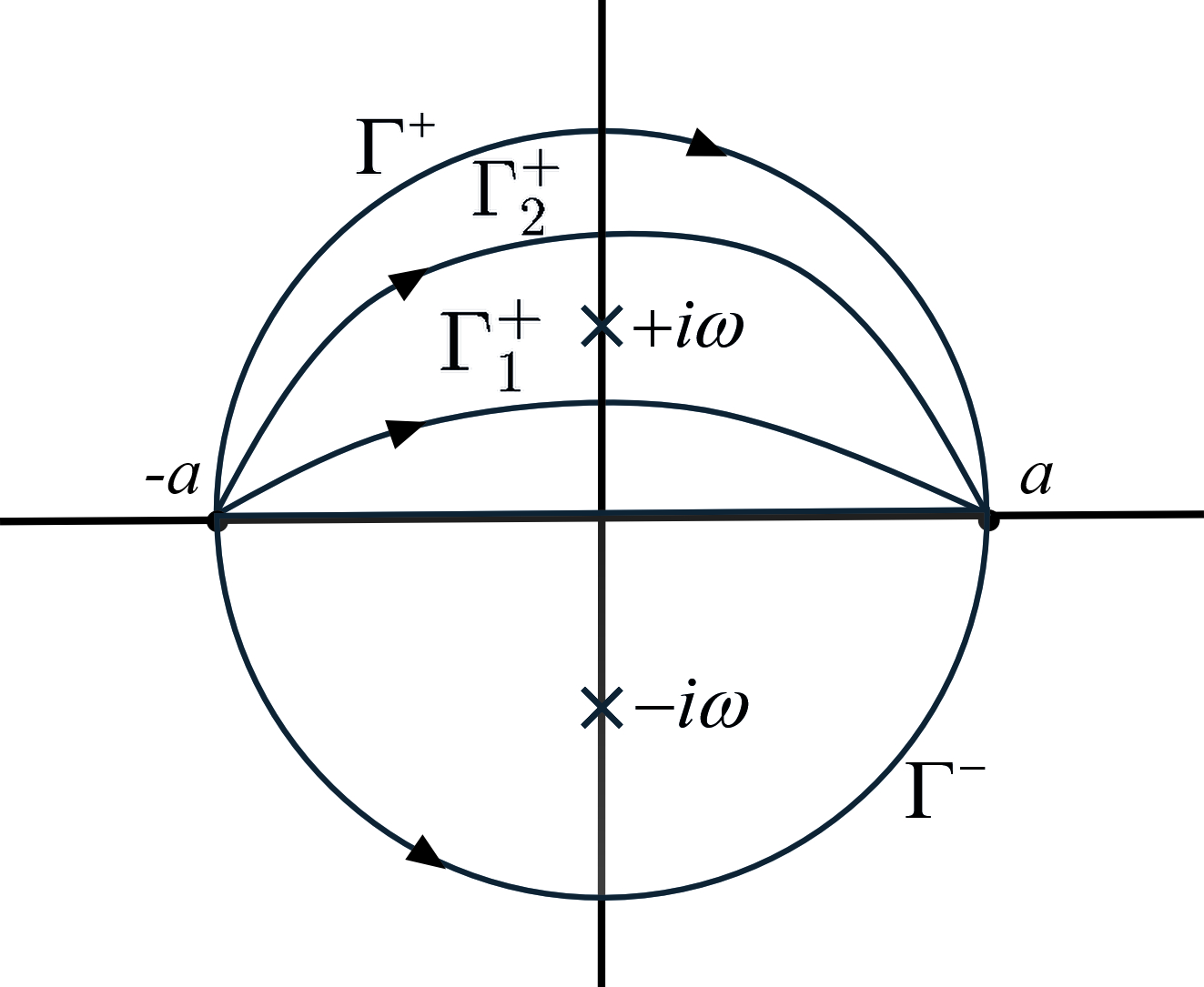}
	\caption{Possible contours of representation of the integral $\int_a^b f(x)(\omega + x)^{-1}\,\mathrm{d}x$. All of them can represent the integral, but not everyone allows term by term integration to be implemented.}
	\label{fig:boat3}
\end{figure}

Finally, let us consider the limit as $a\rightarrow\infty$. Under the condition that the given Stieltjes integral converges absolutely, equation \eqref{lbvrep4} has a well-defined limit as $a\rightarrow\infty$. The limit is the desired value of the Stieltjes integral and is given by
\begin{equation}\label{lbvinfinite}
\int_{-\infty}^{\infty} \frac{f(x)}{\omega^2 + x^2} \, \mathrm{d}x =   \sum_{j=0}^{\infty} (-1)^j \omega^{2j} \, \mathrm{LBV}\!\!\int_{-\infty}^{\infty} \frac{f(x)}{x^{2j+2}}\,\mathrm{d}x + \frac{\pi}{\omega} f(i\omega),\;\;\; \omega> 0,
\end{equation}
with the infinite series converging absolutely for all $\omega>0$. Then we have evaluated the given Stieltjes integral in terms of the lower boundary value of the divergent integrals appearing upon naive term by term integration of the binomially expanded kernel. (A detailed proof of equation \eqref{lbvinfinite} is given in the Supplementary material.)

\subsubsection{Integration by the $\mathrm{UBV}$}
We now evaluate the same integral using the upper boundary value of the divergent integrals. The key step again is to represent the given integral as contour integral using the contour of the UBV contour integral representation. Using the semicircle centered at the origin in the lower complex plane $\Gamma^-$ for the contour in equation \eqref{lowerrep}, the same integral assumes the representation
\begin{equation}
\int_{-a}^{a} \frac{f(x)}{(\omega^2 + x^2)} \, \mathrm{d}x = \int_{\Gamma^-} \frac{f(z)}{\omega^2 + z^2}\, \mathrm{d}z + \frac{\pi}{\omega} f(-i\omega) ,
\end{equation}
where the second term arises from the simple pole at $z=-i\omega$. By the same steps done above with the contour $\Gamma^+$ replaced only with $\Gamma^-$, we obtain the result
\begin{equation}\label{ubvinfinite}
\int_{-\infty}^{\infty} \frac{f(x)}{\omega^2 + x^2} \, \mathrm{d}x =   \sum_{j=0}^{\infty} (-1)^j \omega^{2j} \, \mathrm{UBV}\!\!\int_{-\infty}^{\infty} \frac{f(x)}{x^{2j+2}}\,\mathrm{d}x + \frac{\pi}{\omega} f(-i\omega),\;\;\; \omega> 0,
\end{equation}
The equality of the right hand sides of equations \eqref{lbvinfinite} and \eqref{ubvinfinite} is not apparent but their equality can be established from equation \eqref{bvdifference} and the fact that $f(z)$ has an everywhere convergent expansion about the origin following from the assumption that it is entire. 

\subsubsection{Integration by the $\mathrm{FPI}$}
Finally we consider integration of the given integral by finite parts. Again we need to represent the same integral in terms of the contours $\Gamma^+$ and $\Gamma^-$. The poles are enclosed by the circle formed by these semicircles. Then equation \eqref{fpirep} yields the integral representation
\begin{eqnarray}
\int_{-a}^a \frac{f(x)}{\omega^2 + x^2}\, \mathrm{d}x &=& \frac{1}{2} \left(\int_{\Gamma^+} \frac{f(z)}{\omega^2 + z^2} \mathrm{d}z+\int_{\Gamma^-} \frac{f(z)}{\omega^2 + z^2} \mathrm{d}z  \right) \nonumber \\
&& \hspace{20mm}+ \frac{\pi}{2\omega} \left[f(i\omega)+f(-i\omega)\right],
\end{eqnarray}
where the second term comes from the contributions of the residues at $z=\pm i \omega$. Again we expand the kernel in the same manner that we have done above and proceed similarly. Then we obtain the same integral in terms of the finite part of the divergent integrals
\begin{eqnarray}\label{infinitelimits}
\int_{-\infty}^{\infty} \frac{f(x)}{\omega^2 + x^2} \, \mathrm{d}x &=&  \sum_{j=0}^{\infty} (-1)^j \omega^{2j} \, \mathrm{FPI}\!\!\int_{-\infty}^{\infty} \frac{f(x)}{x^{2j+2}} \,\mathrm{d}x\nonumber \\
&& \hspace{20mm}+ \frac{\pi}{\omega} \left[f(i\omega)+f(-i\omega)\right],\;\;\; \omega> 0, 
\end{eqnarray}
with the infinite series converging absolutely for all $\omega>0$. The equality of this representation with the first two others can be established using the relationship between the boundary values and the finite part integral. 

\subsubsection{When $f(z)$ is not entire} In the above evaluation of the Stieltjes transform, it has been assumed that the complex extension $f(z)$ is entire. Under this condition the resulting infinite series converges absolutely for all values of $\omega>0$ and, by extension, to all complex $\omega$. What happens to the convergence of the infinite series when $f(z)$ has singularities? Foremost, the contour integral representations of the LBV, UBV and FPI require that the path of representations $\gamma^+$ and $\gamma^-$ do not cross any singularity of $f(z)$ when they are deformed to the real line. That is no singularity of $f(z)$ is enclosed by the region bounded by $\gamma^{+}$ and $\gamma^-$. Also $\gamma^{\pm}$ must also be chosen such that the expansion converges uniformly along them to allow the interchange of the summation and the integration. Now if $z_0$ is the closest singularity of $f(z)$ to the origin, then it is necessary that $\omega<|z_0|$; otherwise, we can not have the paths $\gamma^{\pm}$ simultaneously satisfy both conditions. Therefore the evaluations \eqref{lbvinfinite}, \eqref{ubvinfinite} and \eqref{infinitelimits} remain valid provided that $|\omega|<|z_0|$, so that the radius of convergence is determined by the nearest singularity of $f(z)$. 

\section{Term by term integration of the Hilbert transform}\label{hilbertransform}
In the above Stieltjes transform the missing terms arise from the poles of the kernel of the transformation away from the real line which is the entire support of the integral transform. We now give an example when the absence of such poles do not lead to missing terms upon term by term integration with the appropriate choice of the value assigned to the emerging divert integrals. Let us consider the evaluation of the Hilbert transform given by
\begin{equation}
H[f]=\mathrm{PV}\!\!\!\int_{-\infty}^{\infty}\frac{f(x)}{\omega-x}\,\mathrm{d}x ,
\end{equation}
where PV stands for the principal value integral and we have dropped the factor $1/\pi$ in the  standard definition of the Hilbert transform to simplify our notation. Again an attempt to evaluate the integral by expanding the kernel about $\omega  =0$ and integrating term by term the resulting expansion leads to an infinite series of divergent integral due to a non-integrable singularity at the origin. 

We now proceed to use divergent integrals in evaluating the Hilbert transform about $\omega=0$. We assume that $f(x)$ possesses a complex extension $f(z)$ that is analytic in the entire real line $\mathbb{R}$ and at most meromorphic  elsewhere. We use the contour integral representation of the principal value,
\begin{equation}
H[f]=\frac{1}{2}\int_{\gamma^+}\frac{f(z)}{\omega-z}\,\mathrm{d}z + \frac{1}{2}\int_{\gamma^-}\frac{f(z)}{\omega-z}\,\mathrm{d}z 
\end{equation}
which corresponds to $n=0$ in equation \eqref{fpi}. For sufficiently small $\omega$ we expand binomially the kernel,
$(\omega-z)^{-1} = \sum_{k=0}^{\infty} \omega^k z^{-k-1}$. 
Then with the paths chosen to satisfy uniform convergence of the expansion, the expansion can be substituted back and integrated term by term. We have
\begin{equation}
H[f]=\sum_{k=0}^{\infty} \omega^k \frac{1}{2} \left(\int_{\gamma^+}\frac{f(z)}{z^{k+1}}\,\mathrm{d}z+\int_{\gamma^-}\frac{f(z)}{z^{k+1}}\,\mathrm{d}z\right) .\label{mata}
\end{equation}
The radius of convergence of this series is $|z_0|$, where $z_0$ is the singularity of $f(z)$ nearest to the origin, in accordance with our discussion above.

Using the contour  integral representations of the LBV and the UBV, equations \eqref{pos2} and \eqref{neg1}, respectively, and their relationship, given by equation \eqref{bvdifference}, the Hilbert transform \eqref{mata} assumes the evaluations
\begin{equation}
H[f]=\sum_{k=0}^{\infty} \omega^k \; \mathrm{LBV}\!\!\!\int_{-\infty}^{\infty} \frac{f(x)}{x^{k+1}}\,\mathrm{d}x +\pi i f(\omega) ,
\end{equation}
\begin{equation}
H[f]=\sum_{k=0}^{\infty} \omega^k \; \mathrm{UBV}\!\!\!\int_{-\infty}^{\infty} \frac{f(x)}{x^{k+1}}\,\mathrm{d}x -\pi i f(\omega) .\end{equation}
As expected term by term integration using either LBV or UBV leads to a missing term. However, taking the two integrals together in equation \eqref{mata} we find that they evaluate exactly to the FPI of the divergent integral $\int_{-\infty}^{\infty} f(x) x^{-k-1}\,\mathrm{d}x$. Then the Hilbert transform assumes the representation 
\begin{equation}
H[f]=\sum_{k=0}^{\infty} \omega^k\;\; \mathrm{FPI}\!\!\!\int_{-\infty}^{\infty} \frac{f(x)}{x^{k+1}}\, \mathrm{d}x,
\end{equation}
which is just the result of term by term integration using the FPI without any correction term. This a consequence of the absence of poles of the kernel in the complex plane and the exact cancellation of the residues along the real line coming from the two integrals along $\gamma^+$ and $\gamma^-$.  Hence it is not always the case that term by term integration involving divergent integrals will lead to missing terms. 

\section{Calculus of divergent integrals in term by term integration}\label{calculus}
We now formalize the set of rules governing the use of divergent integrals in term by term integration. Suppose we have an absolutely convergent integral
\begin{equation}\label{convergentintegral}
\int_I h(x)\, \mathrm{d}x
\end{equation}
where $I$ is some interval in the real line $\mathbb{R}$. To make the essential features of the calculus of divergent integrals in term by term integration manifest clearly, we make the following simplifying assumptions on the properties of the integrand $h(x)$: (i) the complex extension of $h(x)$, $h(z)$, is analytic in $I$, with at most a finite number of poles elsewhere; (ii)  $h(x)$ admits the expansion $h(x)=\sum_{l=0}^{\infty} h_l(x)$ in some subset $J$ of $I$, where the $h_l(x)$'s posses a common set of non-integrable singularities in the interval $I$; (iii) the complex extension $h(z)$ admits the expansion $h(z)=\sum_{l=0}^{\infty} h_l(z)$ at least along some path in $\mathbb{C}$ which can be continuously deformed to coincide with the interval $I$, and the $h_l(z)$'s are the complex extensions of the $h_l(x)$'s; (iv) the singularities of the $h_l(x)$'s do not occur at infinity. 

Now substituting the expansion back into the integral \eqref{convergentintegral} and performing term by term integration lead to the formal expression
\begin{equation}\label{illdefined}
\sum_{l=0}^{\infty} \int_I h_l(x) \mathrm{d}x
\end{equation} 
which is an infinite series of divergent integrals due to the non-integrable singularities of the $h_k(x)$'s in the interval of integration $I$. To give meaning to expression \eqref{illdefined} by an explicit use of a particular interpretation of the divergent integrals, we follow the following steps which constitute the calculus of divergent integrals arising from term by term integration under the assumptions stated above. \\ 

	 1. {\it Assign values to the divergent integrals, $\int_I h_l(x)\mathrm{d}x$, by deciding how to remove the divergent contributions coming from the offending singularities in the integration.} 	This involves modifying the integrand or the limits of integrations to isolate the singularities in the integral $\int_I h_l(x)\mathrm{d}x$. This will lead to a parametrized sequence of functions involving convergent integrals. Then for every $h_l(x)$ in the expansion, the corresponding value is given by the limiting procedure
	\begin{equation}\label{divergentvalue}
	\#\!\int_I h_l(x) \mathrm{d}x = \lim_{\vec{\alpha}\rightarrow \vec{\alpha}_0} \mathcal{I}_l(\vec{\alpha}), \;\;\; l=0, 1, 2, \dots
	\end{equation}
	where $\vec{\alpha}$ is a set of parameters defining the means of removing the divergences in the integration and $\vec{\alpha_0}$ is the set of values of the parameter $\vec{\alpha}$ that formally recovers the original divergent integral $\int_I h_l(x)\mathrm{d}x$. A fundamental requirement on the value defined by equation \eqref{divergentvalue} is that it must coincide with the value of the integral when its happens to converge absolutely. Moreover, it must admit a contour integral representation as described in the next item.\\
	
	2. {\it Obtain the contour integral representation of the interpretation assigned to the divergent integrals.}
	The contour integral representation is characterized by a set of distinct paths $\{\mathrm{C}_k^{\#}, k=1,...,M\}$ and a set of analytic kernels $\{G_k^{\#}(z),k=1,..,M\}$ such that the value assumes the representation
	\begin{eqnarray}\label{crep}
	\#\!\!\int_I h_l(x)\,\mathrm{d}x = \sum_k \int_{\mathrm{C}_k^{\#}} G_k^{\#}(z) h_l(z)\, \mathrm{d}z        , 
	\end{eqnarray}
for all $l=0,1,2,\dots $. The non-integrable singularities of $h_l(x)$ in the interval $I$ must not lie in any of the paths $\mathrm{C}_k^{\#}$. We take two paths to be distinct if one cannot be continuously deformed to the other without crossing a  singularity of $h_l(z)$. 
	We will refer to the (least) number of contours in the representation as the genus of the value of the divergent integral. 
	
	The values $\mathrm{UBV}\!\!\int_a^b f(x) (x-x_0)^{-n-1}\mathrm{d}x$  and the $\mathrm{LBV}\!\!\int_a^b f(x) (x-x_0)^{-n-1}\mathrm{d}x$ are both genus-1 with analytic kernels $G(z)=1$, and with respective contour of integrations $\gamma^-$ and $\gamma^+$ indicated in Figure-1. On the other hand, the value $\mathrm{FPI}\!\!\int_a^b f(x) (x-x_0)^{-n-1}\mathrm{d}x$ is genus-2 with kernels $G_k(z)=1/2$, $k=1,2$, and contours of integrations $\gamma^+$ and $\gamma^-$. In \cite{galapon2}, we have the following contour integral representations of a couple of finite part integrals
	\begin{equation}
	\mathrm{FPI}\!\!\int_0^a \frac{f(x)}{x^m}\mathrm{d}x = \frac{1}{2\pi i} \int_{\mathrm{C}} \frac{f(z)}{z^m}\left(\log z - i\pi \right)\mathrm{d}z, 
	\end{equation}
\begin{equation}
\mathrm{FPI}\!\!\int_0^a \frac{f(x)}{x^{m+\nu}}\,\mathrm{d}x = \frac{1}{(e^{-i 2\pi \nu }-1)} \int_{\mathrm{C}} \frac{f(z)}{z^{m+\nu}}\,\mathrm{d}z
\end{equation}
	where $m=1, 2, \dots$, $0<\nu<1$, and $\mathrm{C}$ is the path that straddles the branch cut of $\log z$ which is the positive real axis, and that starts and ends at the point $a$. Both are of genus-1 with respective kernels $(\log z - i\pi)/2\pi i$ and $1/(e^{-i 2\pi \nu}-1)$.\\

	3. {\it Represent the given convergent integral $\int_I h(x)\mathrm{d}x$ as a contour integral up to a term using a set of paths $\{\tilde{C}_k^{\#}\}$ homotopic to the representation paths $\{\mathrm{C}_k^{\#}\}$ of the chosen interpretation of the divergent integrals.} 
This requires deciding on the complex extension, $h(z)$, of $h(x)$. The choice for $h(z)$ must be consistent with the chosen interpretation of the divergent integrals arising from term by term integration. The representation is obtained by considering the contour integral 
	$\sum_k \int_{\tilde{C}_k^{\#}} G_k^{\#}(z) h(z)\, \mathrm{d}z$,
	where none of the singularities of $h(z)$ lie along any of the $\tilde{C}_k^{\#}$'s. Then the paths, $\tilde{C}_k^{\#}$'s, are deformed continuously to the interval $I$ to recover the given integral in the form 
	\begin{equation}\label{form}
	\int_I h(x)\mathrm{d}x = \sum_k \int_{\tilde{C}_k^{\#}} G_k^{\#}(z) h(z)\, \mathrm{d}z + \Delta^{\#},
	\end{equation}
	where the term $\Delta^{\#}$ emerges from the singularities of $h(z)$ crossed by the contours $\tilde{C}_k^{\#}$ in the process of deforming them to the interval $I$. We say that a singularity $z_0$ of $h(z)$ is enclosed by the paths $\{\tilde{C}_k^{\#}\}$ if $z_0$ is crossed by at least by one path $\tilde{C}_K^{\#}\in\{\tilde{C}_k^{\#}\}$ when the $\tilde{C}_k^{\#}$'s are deformed into $I$. The term $\#\Delta$ may be present or absent depending on whether a singularity of $h(z)$ is enclosed by the paths $\tilde{C}_k^{\#}$ or not. With the intention to use the representation \eqref{form} to implement term by term integration, one may initially use the contours that can be consistently constructed which enclose the maximum number of singularities of $h(z)$. \\
	
	4. {\it Implement term by term integration. } In order for the term by term integration to be valid, the contours must be chosen such that  the expansion $h(z)=\sum_{l=0}^{\infty} h_l(z)$ converges uniformly for all $z\in\tilde{C}_k^{\#}$, for every $k=1,\dots,M$. Under this condition, the expansion for $h(z)$ can be substituted back into equation \eqref{form} and integrated term by term to yield
\begin{equation}
\int_Ih(x)\mathrm{d}x = \sum_{l=0}^{\infty} \sum_{k=1}^M \int_{\tilde{C}^{\#}_k} G_k(z)h_l(z)\mathrm{d}z + \Delta^{\#} .
\end{equation} 
While the paths $\{\tilde{C}_k^{\#}\}$ are homotopic to the representation paths $\{\mathrm{C}_k^{\#}\}$, it is not necessary that the contour integrals in the first term are the respective values $\#\int_I h_l(x)\mathrm{d}x$ of the divergent integrals $\int_I h_l(x)\mathrm{d}x$. For the integrals to be identified as the values $\#\int_I h_l(x)\mathrm{d}x$, it is necessary that the only singularities of $h_l(z)$ enclosed by the contours $\tilde{C}_k^{\#}$ are those enclosed in the representation \eqref{crep}. If this condition is satisfied, the contour integrals in the series reduces to the chosen value of the divergent integrals, so that we have the evaluation
	\begin{equation}\label{finalvalue}
	\int_I h(x)\mathrm{d}x = \sum_{l=0}^{\infty} \#\!\!\!\int_{I} h_l(x)\, \mathrm{d}x + \Delta^{\#} .
	\end{equation} 	
The first term represents the contributions coming from naive term-by-term integration of the expanded integrand with the chosen value $\#$ assigned to the divergent integrals; the second term represents the terms missed out by naive term-by-term integration corresponding to the value $\#$ arising from the singularities of $h(z)$ . Not all singularities of $h(z)$ contribute in $\#\Delta$ but only those that are inevitably enclosed by the contours $\{\tilde{C}_k^{\#}\}$ when uniform convergence of the expansion is enforced along each contour $\tilde{C}_k^{\#}$. \\ 

In equation \eqref{finalvalue}, it is assumed that there are infinite number of terms in the expansion. What happens when there are only a finite number of terms? As discussed above, the missed out term $\Delta^{\#}$ arises from the singularities of the integrand in the complex plane. This contribution cannot be avoided by some choice of the contours of integration when uniform convergence is demanded for the interchange of integration and infinite summation to be valid. However, the situation is different when the expansion involves only a finite number of terms because the integration can be distributed term by term without satisfying uniformity condition. This allows us to chose the contours of the representation of a given integral that avoid all the singularities of $h(z)$ so that we have $\Delta^{\#}=0$ in equation \eqref{form}. Assuming all the above conditions on the function $h(x)$ except that it has the finite expansion $h(x)=\sum_{k=0}^{N}h_k(x)$, then we have the result
\begin{equation}\label{finite}
\int_I h(x)\,\mathrm{d}x = \sum_{k=0}^N \#\!\!\int_I h_k(x)\,\mathrm{d}x ,
\end{equation} 
provided the assigned value $\#$ to the divergent integrals $\int_I h_k(x)\mathrm{d}x$ admits a contour integral representation described above.  In this case, naive term by term integration works without missing any contribution from the singularities of the integrand in the complex plane. This is always true when there are only a finite number of terms. The case of the FPI in the Hilbert transform is an exception to the general rule of the presence of missing terms when there are infinite number of terms in the expansion.

Clearly the calculus does not prescribe a specific interpretation to the divergent integrals arising from term by term integration but allows us to assign any values to them, with the missing terms automatically emerging from the chosen interpretation of the divergent integrals itself. By ``any value'' we mean any interpretation of the divergent integrals that admits a contour integral representation in the form indicated above. 

\section{An application}\label{application}
Let us give an application of the calculus of divergent integrals developed here that goes beyond the Stieltjes transform and its cousin Hilbert transform. The LBV, UBV and FPI are relevant in the evaluation of the Fourier transform of product of exponential type functions. Examples of such Fourier transforms are given by
\begin{equation}\label{fourier}
\int_{-\infty}^{\infty} e^{i a x} \prod_{k=1}^N \frac{\sin^{n_k}b_k(x-c_k)}{(x-c_k)^{n_k}}
\end{equation}
for positive integers $N$ and $n_k$, and reals $a$, $b_k$ and $c_k$. These integrals appear in a quantum measurement model of decoherence without the aid of an environment \cite{galaponepl}. Exact evaluation of the integrals is necessary to obtain the exact dynamics of the occurence of decoherence. However, we will not do the physical analysis here. It is enough for us to outline how the calculus of divergent integrals developed here allows exact evaluation of the Fourier transform. 

The idea is to induce divergence in the otherwise convergent integral by expanding the product followed by term by term integration. We illustrate the method by evaluating the integral
\begin{equation}\label{given}
\int_{-\infty}^{\infty} \frac{\sin^2(x)}{x^2}\,\mathrm{d}x = \pi .
\end{equation}
To introduce divergent integrals in this convergent integral, we make the substitution $\sin x = (e^{ix}-e^{i x})/2i$ and perform term by term integration. The result is a sum of divergent integrals,
\begin{equation}
-\frac{1}{4}\int_{-\infty}^{\infty} \frac{e^{2 i x}}{x^2}\, \mathrm{d}x+\frac{1}{2}\int_{-\infty}^{\infty} \frac{\mathrm{d}x}{x^2}-\frac{1}{4}\int_{-\infty}^{\infty} \frac{e^{-2 i x}}{x^2}\, \mathrm{d}x .
\end{equation}
Since there are only a finite number of terms in the summation, then, according to our results above, the divergent integrals can be assigned any value provided the value admits a contour integral representation. Then the integral evaluates to
\begin{equation}
\int_{-\infty}^{\infty} \frac{\sin^2(x)}{x^2}\,\mathrm{d}x=-\frac{1}{4}\#\!\!\!\int_{-\infty}^{\infty} \frac{e^{2 i x}}{x^2}\, \mathrm{d}x+\frac{1}{2}\#\!\!\!\int_{-\infty}^{\infty} \frac{\mathrm{d}x}{x^2}-\frac{1}{4}\#\!\!\!\int_{-\infty}^{\infty} \frac{e^{-2 i x}}{x^2}\, \mathrm{d}x ,\label{pepe}
\end{equation}
where $\#$ identifies the assigned value to the divergent integrals. 

We use the LBV, UBV and the FPI of the divergent integrals in expression \eqref{pepe} to evaluate the given integral \eqref{given}. We give the values relevant not only in the evaluation of integral \eqref{given} but also in the evaluation of the entire class of Fourier transform in \eqref{fourier}. 
The values are given by
\begin{equation}\label{valuelbv}
\mathrm{LBV}\!\!\!\int_{-\infty}^{\infty} \frac{e^{i \sigma x}}{(x-x_0)^n}= -\frac{2 \pi i^n  }{(n-1)!} \sigma^{n-1} e^{i \sigma x_0} H(-\sigma),\;\; \mathrm{LBV}\!\!\int_{-\infty}^{\infty}\frac{\mathrm{d}x}{(x-x_0)^n}=-i\pi \delta_{n1}
\end{equation}
\begin{equation}\label{valueubv}
\mathrm{UBV}\!\!\!\int_{-\infty}^{\infty} \frac{e^{i \sigma x}}{(x-x_0)^n}= \frac{2 \pi i^n}{(n-1)!} \sigma^{n-1} e^{i \sigma x_0} H(\sigma),\;\; \mathrm{UBV}\!\!\int_{-\infty}^{\infty}\frac{\mathrm{d}x}{(x-x_0)^n}=i\pi \delta_{n1}
\end{equation}
\begin{equation}\label{valuefpi}
\mathrm{FPI}\!\!\!\int_{-\infty}^{\infty} \frac{e^{i \sigma x}}{(x-x_0)^n}=  \frac{\pi i^n }{(n-1)!} \sigma^{n-1} e^{i \sigma x_0}\,\mathrm{sgn}(\sigma),\;\; \mathrm{FPI}\!\!\!\int_{-\infty}^{\infty}\frac{\mathrm{d}x}{(x-x_0)^n}=0
\end{equation}
for $n=1, 2, 3, \dots$, where $H(x)$ is the Heaviside step function and $\mathrm{sgn}(x)$ is the sign function. These values are obtained by an explicit evaluation of the contour integral representations of the LBV, UBV and FPI using the calculus of residues. The contour integrals along $\gamma^{\pm}$ are obtained by closing the contours by a great semi-circle in the upper or lower half of the complex plane, depending on the sign of the parameter $\sigma$ when $\sigma\neq $ or whichever when $\sigma=0$. 

When the relevant values of the parameters are substituted in equations \eqref{valuelbv}, \eqref{valueubv} and \eqref{valuefpi} for the divergent integrals in equation \eqref{pepe}, we obtain the value $\pi$, regardless which value is used (LBV, UBV or FPI). 

\section{Conclusion}\label{conclusion}
We have seen how divergent integrals arise from a well-defined convergent integral from an interchange of the order of integration and summation when the required uniformity for the interchange to be valid is not satisfied. However, the expansion of the integrand and the eventual distribution of integration typically does not arise from a blind attempt to evaluate an integral, but may in fact be physically motivated. For example, the effective diffusivity by passive advection in laminar and turbulent flows assumes a representation which involves the generalized Stieltjes transform $\int_{-\infty}^{\infty}f(x) (\omega^2 + x^2)^{-1}\mathrm{d}x$. It is desired there to obtain the behavior of the effective diffusivity for large Peclet number which corresponds to small $\omega$ in the Stieltjes transform. Then one maybe motivated to expand the kernel about $\omega=0$ and then perform a term by term integration. But, as we have seen here, this leads to an infinite series of divergent integrals due to the non-uniform convergence of the expansion in the interval of integration. Because of the divergence in each term, one may consider the attempt as a failure and abandon the approach. 

But here we have shown that the divergent integrals can be tackled so that the attempt at an expansion about small $\omega$ can be made meaningful by using the calculus of divergent integrals introduced here. The calculus indicates that the divergent integrals are not pathologies that have to be avoided at all cost but are in fact harbingers of the presence of contributions lurking in the complex plane. The calculus allows us to assign arbitrary values to the divergent integrals and automatically make the necessary corrections by picking up the lurking contributions from the complex singularities of the integrand. Since the calculus does not prescribe a specific interpretation of the divergent integrals, it offers flexibility in its implementation in applications---one may choose the most convenient interpretation that can be ascribed to the divergent integrals. Overall the results reported here show that divergent integrals are not as ill-defined as their divergence suggests but are in fact meaningful mathematical objects that can be used to solve well-defined problems.

\section*{Acknowledgment}
The author acknowledges the Office of the Chancellor of the University of the Philippines Diliman, through the Office of the Vice Chancellor for Research and Development, for funding support through the Outright Research Grant Project No. 171711 PNSE.

\renewcommand{\theequation}{S-\arabic{equation}}
\setcounter{equation}{0}  
\setcounter{section}{0}

\section*{Supplement}
\subsection*{Proof of equation (44)}
Here we give an independent proof of  equation (44) in the main text by using the contour $\Gamma^+$ in Figure-4 of this Supplement and by means of the expansion of the kernel in the form
\begin{equation}
\frac{1}{\omega^2+z^2} = \sum_{j=0}^{n-1}(-1)^j \frac{\omega^{2j}}{z^{2j+2}} + (-1)^n \frac{\omega^{2n}}{z^{2n} (\omega^2 + z^2)}, \;\; n=1, 2, \dots .
\end{equation}
Substituting this expansion back into the contour integral representation 
\begin{equation}\label{lbvrep}
\int_{-a}^{a} \frac{f(x)}{(\omega^2 + x^2)} \, \mathrm{d}x = \int_{\Gamma_2^+} \frac{f(z)}{\omega^2 + z^2}\, \mathrm{d}z + \frac{\pi}{\omega} f(i\omega),
\end{equation}
yields
\begin{eqnarray}\label{lbvrep2}
\int_{-a}^{a} \frac{f(x)}{(\omega^2 + x^2)} \, \mathrm{d}x &=& \sum_{j=0}^{n-1}(-1)^j \omega^{2n}\int_{\Gamma^+} \frac{f(z)}{z^{2j+2}}\, \mathrm{d}z \nonumber \\
&&\hspace{12mm}+ (-1)^n \omega^{2n} \int_{\Gamma^+} \frac{f(z)}{z^{2n} (\omega^2 + z^2)}\, \mathrm{d}z + \frac{\pi}{\omega} f(i\omega) .
\end{eqnarray}
Then we have 
\begin{equation}\label{lbvrep3}
\int_{-a}^{a} \frac{f(x)}{(\omega^2 + x^2)} \, \mathrm{d}x = \sum_{j=0}^{n-1}(-1)^j \omega^{2n} \mathrm{LBV}\!\!\int_{-a}^a \frac{f(x)}{x^{2j+2}}\, \mathrm{d}x +  R_n(\omega) + \frac{\pi}{\omega} f(i\omega) .
\end{equation}
where the remainder term $R_n(\omega)$ is given by
\begin{equation}
(-1)^n \omega^{2n} \int_{\Gamma^+} \frac{f(z)}{z^{2n} (\omega^2 + z^2)}\, \mathrm{d}z . 
\end{equation}

We wish now to establish that the limit of the remainder term vanishes in the limit as $n\rightarrow\infty$ provided $\omega<a$. We find a bound for the remainder term by using the parametrization $z=a e^{i\theta}$ along the semicircular path of integration. We obtain
\begin{eqnarray}
\left|R_n(\omega)\right| \leq \left(\frac{\omega}{a}\right)^{2n} M(a) ,
\end{eqnarray}where
\begin{equation}\label{ma}
M(a)=\frac{1}{2} \left(\int_{\pi}^0 \frac{|f(a e^{i\theta})|}{|\omega^2 + a^2 e^{i2 \theta}|}\,\mathrm{d}\theta +\int_{-\pi}^0 \frac{|f(a e^{i\theta})|}{|\omega^2 + a^2 e^{i2 \theta}|}\,\mathrm{d}\theta\right)
\end{equation} 
The quantity in the parenthesis is bounded and independent of $n$. Then the reminder term vanishes in the limit as $n\rightarrow\infty$ provided $\omega<a$, so that the integral evaluates to
\begin{equation}\label{lbvrep33}
\int_{-a}^{a} \frac{f(x)}{(\omega^2 + x^2)} \, \mathrm{d}x = \sum_{j=0}^{\infty}(-1)^j \omega^{2n} \mathrm{LBV}\!\!\int_{-a}^a \frac{f(x)}{x^{2j+2}}\, \mathrm{d}x +\frac{\pi}{\omega} f(i\omega) .
\end{equation}
which is just equation (44).

\subsection*{Proof of equation (45)}
\begin{figure}
	\includegraphics[scale=0.6]{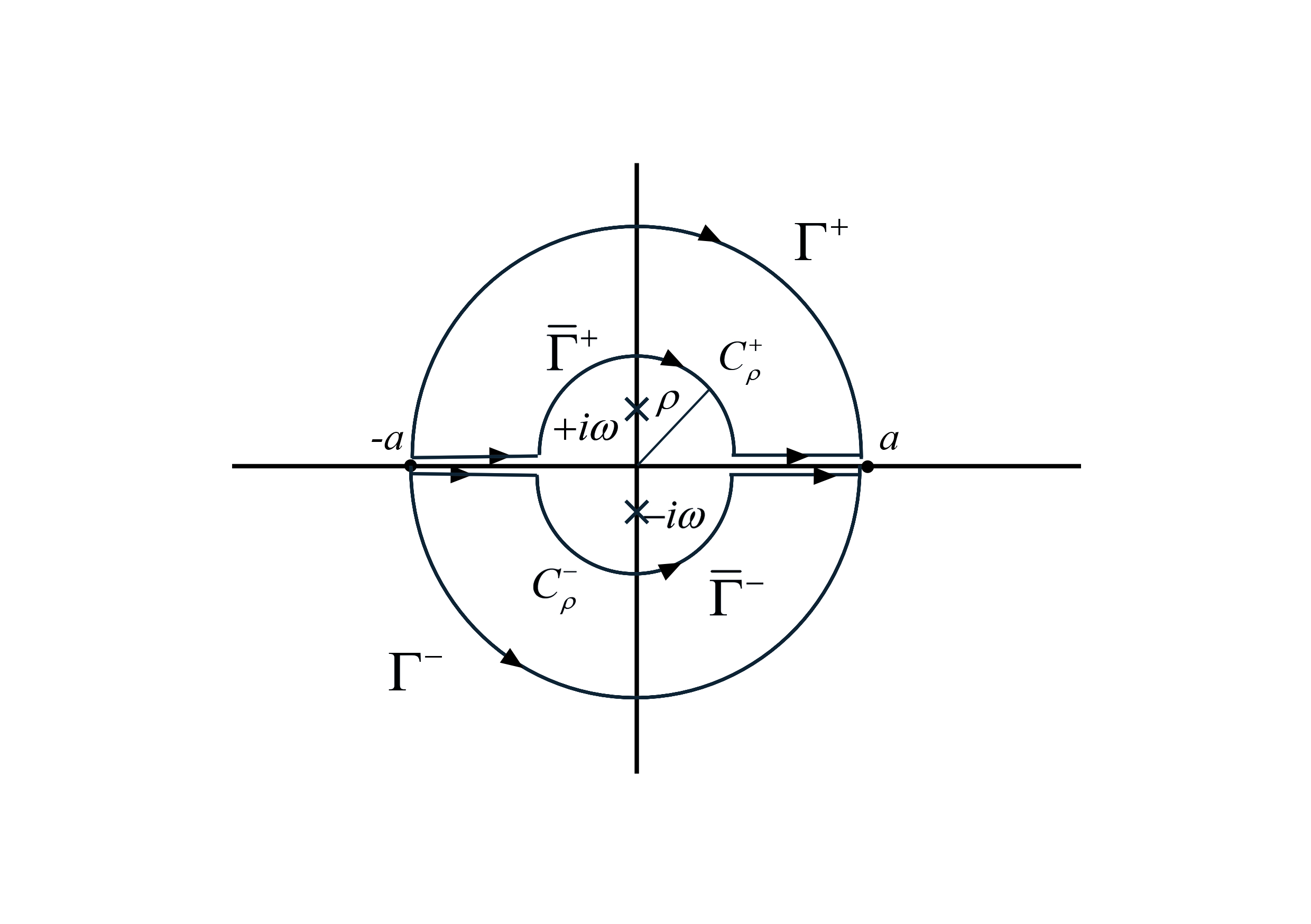}
	\caption{The deformation of the contours of integrations $\Gamma^+$ and $\Gamma^-$.}
	\label{fig:boat3}
\end{figure}
We now establish equation (45) in the main text. From the above condition, the infinite series in equation \eqref{lbvrep33} converges absolutely for all $\omega<a$. Under such condition, there always exists some fixed $\rho$ satisfying $\omega<\rho<a$. We deform the contour of integration $\Gamma^+$ into the contour $\bar{\Gamma}^+$ as depicted in Figure-3 so that the LBV assumes the form
\begin{eqnarray}\label{intbylbv}
\mathrm{LBV}\!\!\int_{-a}^{a} \frac{f(x)}{x^{2j+1}} \, \mathrm{d}x  =   \int_{-a}^{-\rho} \frac{f(x)}{x^{2j+1}}\, \mathrm{d}x  + \int_{C_{\rho}^+}\frac{f(z)}{z^{2j+1}}\mathrm{d}z 
+\int_{\rho}^{a} \frac{f(x)}{x^{2j+1}}\, \mathrm{d}x  .
\end{eqnarray}
By assumption the integral $\int_{-\infty}^{\infty}f(x)(\omega^2 + x^2)^{-1} \, \mathrm{d}x$ converges so that $f(x)/x^2 = o(1/x)$ as $|x|\rightarrow \infty$. Then
\begin{eqnarray}
\left|\int_{\rho}^{a} \frac{f(x)}{x^{2j+1}}\,\mathrm{d}x\right| &\leq& \int_{\rho}^{a} \frac{|f(x)|}{x^{2j+1}} \, \mathrm{d}x \nonumber \\
&\leq & \frac{1}{\rho^{2j-1}} \int_{\rho}^{a} \frac{|f(x)|}{x^2} \, \mathrm{d}x\nonumber \\
&\leq & \frac{1}{\rho^{2j-1}} \int_{\rho}^{\infty} \frac{|f(x)|}{x^2} \, \mathrm{d}x
\end{eqnarray}
Similarly we have the bound
\begin{eqnarray}
\left|\int_{-a}^{-\rho} \frac{f(x)}{x^{2j+1}}\,\mathrm{d}x\right| 
\leq  \frac{1}{\rho^{2j-1}} \int_{\rho}^{\infty} \frac{|f(-x)|}{x^2} \, \mathrm{d}x .
\end{eqnarray}
Moreover, we have the following bound for the integral along the semi-circle $C_{\rho}^+$, 
\begin{equation}\label{ma1}
\left|\int_{C_{\rho}^+}f(z) z^{-2j-1}\mathrm{d}z\right|\leq M(\rho)=\frac{1}{2} \left(\int_{\pi}^0 \frac{|f(\rho e^{i\theta})|}{|\omega^2 + \rho^2 e^{i2 \theta}|}\,\mathrm{d}\theta +\int_{-\pi}^0 \frac{|f(\rho e^{i\theta})|}{|\omega^2 + \rho^2 e^{i2 \theta}|}\,\mathrm{d}\theta\right)
\end{equation} 

Combining all the bounds of the three terms of equation \eqref{intbylbv} we obtain the bound for the lower boundary values of the divergent integrals
\begin{equation}\label{bound}
\left|\mathrm{LBV}\!\!\int_{-a}^{a} \frac{f(x)}{x^{2j+1}}\mathrm{d}x\right|\leq \frac{1}{\rho^{2 j}} \left(M(\rho) + \frac{\rho}{2}\int_{\rho}^{\infty} \frac{\left(|f(x)| + |f(-x)|\right)}{x^{2}}\,\mathrm{d}x\right)<\infty, \;\; \rho< a .
\end{equation}
Consequently we have the bound on the sum
\begin{eqnarray}\label{bound2}
\left|\sum_{j=0}^{\infty}(-1)^j \omega^{2j} \mathrm{LBV}\!\!\int_{-a}^{a} \frac{f(x)}{x^{2j+1}}\,\mathrm{d}x\right| &\leq & \sum_{j=0}^{\infty}\omega^{2j}  \left|\mathrm{LBV}\!\!\int_{-a}^{a}\frac{f(x)}{x^{2j+1}}\, \mathrm{d}x\right| \nonumber \\
&\leq &\sum_{j=0}^{\infty} \frac{\omega^{2j}}{\rho^{2 j}} \left(M(\rho) + \frac{\rho}{2}\int_{\rho}^{\infty} \frac{\left(|f(x)| + |f(-x)|\right)}{x^{2}}\,\mathrm{d}x\right)\nonumber \\
&=& \frac{\rho^2}{\rho^2-\omega^2} \left(M(\rho) + \frac{\rho}{2}\int_{\rho}^{\infty} \frac{\left(|f(x)| + |f(-x)|\right)}{x^{2}}\,\mathrm{d}x\right),
\end{eqnarray}
where the last line follows from the condition that $\omega<\rho$. 

Now let us consider the limit as $a\rightarrow\infty$. Since the right hand side of inequality \eqref{bound} is finite and independent of $a$ for all $a>\rho$, the limit of the LBV  as $a\rightarrow\infty$ exists, so that we have the LBV of the divergent integral $\int_{-\infty}^{\infty} f(x)x^{-2j-1}\mathrm{d}x$,
\begin{equation}
\mathrm{LBV}\!\!\int_{-\infty}^{\infty} \frac{f(x)}{x^{2j+1}}\mathrm{d}x=\lim_{a\rightarrow \infty}\mathrm{LBV}\!\!\int_{-a}^{a} \frac{f(x)}{x^{2j+1}}\mathrm{d}x .
\end{equation} 
Also since the right hand side of the inequality \eqref{bound2} is independent of $a$, the infinite series in the left hand side converges uniformly for all $a>\omega$. This allows us to interchange the order of limit and summation so that we obtain
\begin{equation}\label{lbvinfinite}
\int_{-\infty}^{\infty} \frac{f(x)}{\omega^2 + x^2} \, \mathrm{d}x =   \sum_{j=0}^{\infty} (-1)^j \omega^{2j} \, \mathrm{LBV}\!\!\int_{-\infty}^{\infty} \frac{f(x)}{x^{2j+2}}\,\mathrm{d}x + \frac{\pi}{\omega} f(i\omega),\;\;\; \omega> 0,
\end{equation}
with the infinite series converging absolutely for all $\omega>0$. Then we have evaluated the given convergent Stieltjes integral in terms of the lower boundary value of the divergent integrals appearing upon naive term by term integration of the binomially expanded kernel.

\end{document}